\documentclass[doublecol,linenumbers]{epl2} 

\usepackage{amsmath}
\usepackage[colorlinks,citecolor=blue,linkcolor=blue]{hyperref}

\title{Scattering of exchange spin waves from regions of  modulated magnetization}
\shorttitle{Scattering of exchange spin waves from regions of  modulated magnetization} 

\author{P. Borys\inst{1} \and N. Qureshi\inst{1} \and C. Ordoñez-Romero\inst{2}  \and O. Kolokoltsev\inst{1}}
\shortauthor{P. Borys \etal}

\institute{                    
  \inst{1} Instituto de Ciencias Aplicadas y Tecnolog\'ia, Universidad Nacional Aut\'onoma de M\'exico, Mexico City 04510, Mexico\\
  \inst{2} Instituto de F\'isica, Universidad Nacional Autónoma de México, IFUNAM, Mexico City 04510,
  Mexico
}

\abstract{We investigate the reflection coefficient of spin waves propagating in an ultra-thin ferromagnetic film with regions where saturation magnetization is modulated. We find analytically and using micromagnetic simulations that there are transmission resonances that depend on the width of the regions and on the energy of excitation. Our results resemble the quantum mechanical Ramsauer-Townsend effect in which an electron with certain energies can propagate above a potential field without scattering. Our findings are useful for reconfigurable magnonic devices where the saturation magnetization can be dynamically controlled via a thermal landscape. }

\begin{document}

\maketitle

\section{Introduction }

Spin wave excitations (magnons) represent a Joule-Heat free alternative to traditional semiconductor electronics. Possible applications for spin waves are data storage, filtering, and processing~\cite{Stamps_2014,Sander_2017}. The field of magnonics has had an immense growth in recent years because of technological advances in nanofabrication resulting in miniaturization and metamaterials or artificial crystals   with continuously nonuniform magnetic parameters\cite{Smith2004,serga2010yig,lenk2011building,pirro2014spin}. Even a new term has been coined to describe spin wave propagation in this novel media -- \textit{graded-index magnonics}\cite{Davies_2017,davies2015towards,davies2015graded}. Furthermore, dynamic control over the magnetic parameters has been achieved resulting in fully reconfigurable magnonic devices. For example, by varying the applied magnetic field it has been possible to tune the frequency gaps in a magnonic crystal with possible applications as a spin wave filter \cite{topp_making_2010}. Another example is saturation magnetization being continuously modulated by creating a thermal landscape \cite{kolokoltsev_hot_2012} resulting in control of spin-wave propagation\cite{Vogel2018} and in a fully tunable magnonic crystal\cite{Vogel2015}.

It is important, then, to investigate the phenomena arising from systems where magnetic parameters are modulated. In this study, we investigate how a piece-wise modulation of the saturation magnetization affects the reflection coefficient of propagating exchange spin waves. We find that, depending on the width of the modulated saturation magnetization, the reflection coefficient exhibit resonances.


Our results resemble what happens in the quantum mechanical Ramsauer-Townsend effect (RT)~\cite{Brode1933,Kukolich1968,Grysinski}. A matter wave describing a quantum particle (e.g. an electron) show transmission resonances while propagating through a potential field depending on its energy. In its simplified, one-dimensional form, the RT effect is described as a matter wave  with energy, $E$, propagating above a potential well  of depth  $U$. The RT effect is used to explain why, for some energies, 	

\begin{align}\label{eq:RTresonances}
E=U+\frac{n^2\pi^2}{2mL^2},
\end{align}

electrons of mass $m$ are not scattered when colliding with noble gas atoms of diameter $L$ .  

That spin waves and matter waves exhibit similar features has been known for a long time ~\cite{Schlmann1964,SchlomannBook}. In fact, the equation of motion for the magnetic moments, the Landau-Lifshtz equation, can be written as the stationary  Schr\"odinger equation if magnetic anisotropies and the  dipole interaction are neglected. A spin wave in a ferromagnet magnetized along the $x$ direction, $m_y+i\,m_z\propto\exp(i\omega t)$,  plays the role of a wave function, $\psi$, and hence the squared amplitude $|m_y+i\,m_z|^2$ can be identified with the probability density, $|\psi|^2$. The dispersion relation of a spin wave, approximated as,  

\begin{align}\label{eq:swdisp}
	\omega\simeq\frac{2\gamma A}{M_s}k^2+\gamma\mu_0(H_0+\frac{1}{2}M_s),
\end{align} 

where $A$ is the exchange constant, $\gamma$ is the gyromagnetic ratio, $H_0$ an external field, and $M_s$ the saturation magnetization has the same mathematical form as the dispersion of a quantum mechanical particle,

\begin{align}\label{eq:qmdisp}
	E=\frac{\hbar^2}{2m}k^2+U,
\end{align} 

if the long range dipole energy is approximated as a local interaction as is usual for ultra thin-films, and if ellipticity is neglected. We can, therefore, identify the magnon mass $m^*=\hbar M_s/(4\gamma A)$ and the potential energy $U=\hbar\gamma\mu_0(H_0+\frac{1}{2} M_s)$.



The case where the external field is controlled to act as the potential has been treated before, and tunneling has been observed~\cite{Demokritov2004}. We, on the other hand, control the saturation magnetization. 
\section{Methods}

\begin{figure}
	\includegraphics[width=3.2in]{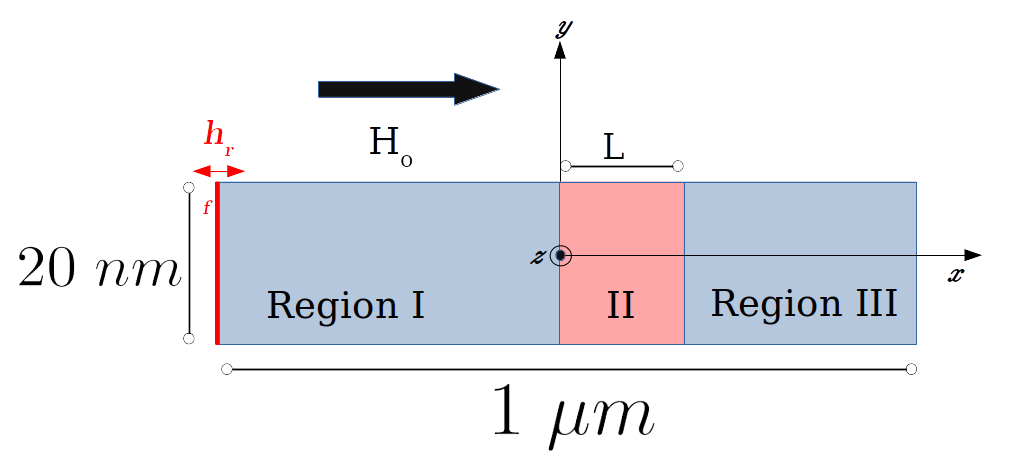}
	\caption{Geometry for spin wave propagation through an ultra-thin magnetic film, where a radio frequency antenna generating an alternating field $h_{rf}$ excites spin waves that propagate along the $x$. The film is magnetized in the $x$ direction. All magnetic parameters are the same in the three regions except the saturation magnetization in Region II. Region II is also characterized by its width, $L$. }	
	\label{fig:geo}	
\end{figure}

We perform micromagnetic simulations using Mumax3~\cite{Vansteenkiste2014} on  a ferromagnetic ultra-thin film $1$ $\mu m$ $\times$ $20$ nm $\times$ $1$ nm discretized using $500$ $\times$ $10$ $\times$ $1$ finite difference cells. The film is magnetized along the long, $x$ axis. A bias field $\mu_0\,H_0=1$ T is also applied in the $x$ direction. The film is characterized by three regions as indicated in fig (\ref{fig:geo}). Exchange constant, $A=3.6$ pJ/m,  is the same in the three regions and typical value for YIG \cite{klingler2014measurements}, while the saturation magnetization, $M_{s}'=100$ kA/m,  in region II is different to the saturation magnetization, $M_s=300$ kA/m in regions I and III. The minimum exchange length $\lambda_{ex}=\sqrt{\frac{2A}{\mu_0 M_s^2}}\approx8$ nm corresponding to $M_s$ in regions I, III is larger than the cell size (2 nm) so that effects of the exchange interaction are predominant. We excite spin waves for $3$ ns on the left extreme of the film by applying a sinusoidal magnetic field, $h_{rf}$, of amplitude $50$ mT and frequency, $\nu=80$ GHz in the $z$ direction spatially localized in the $x$ direction to one cell of the simulation grid.  

\begin{figure}
	\includegraphics[width=3.4in]{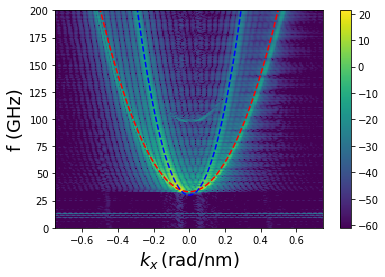}
	\caption{ Contour plot of $m_y(f_i,k_j)$ for $L=200$ nm. In order to excite a large range of frequencies, the exciting transient field was a spatially localized sinc pulse with a cut-off frequency $f_c=500$ GHz.  The color bar indicates the amplitude of the magnetization component in decibels. The red  and blue dashed lines depict the dispersion relation for $M_s=300$ and $M_s'=100$ kA/m, respectively, under the exchange approximation. Simulation agrees well with the theoretical dispersion relations.  }
	\label{fig:DispRel.}		
\end{figure}

We record the spatial profile of dynamic components of the reduced magnetization vector every $1$ ps. We have a matrix $m_{y,z}(t_i,x_j)$. It is possible to obtain the dispersion relation by performing a 2D Fourier Transform on $m_{y,z}(t_i,x_j)$ to get $m_{y,z}(f_i,k_j)$ as shown in figure \ref{fig:DispRel.} where we changed the excitation to a sinc-type pulse to get a broader range of frequencies. To calculate the reflection coefficient, we follow the procedure described in Refs. \cite{dvornik_micromagnetic_2011,dvornik2013micromagnetic,dvornik2013_paper} and \cite{Venkat}. We focus on Region I and get $m_{(y,z),I}(f_i,k_j)$ from where we extract the complex amplitudes $m_{y,I}(80 GHz,\pm k(80 GHz))$ and $m_{z,I}(80 GHz,\pm k(80 GHz))$. Finally, the reflection coefficient is calculated as  

\begin{align}\label{eq:RefExperiment}
	R= \frac{m_{y,I}(-)^2+m_{z,I}(-)^2}{(m_{y,I}(+)^2+m_{z,I}(+))_C}
\end{align}

Where $m_{y,I}$, and $m_{z,I}$ are the complex amplitudes at $80$ GHZ, the plus $(+)$ and $(-)$ signs indicate that we take the $k>0$ or the $k<0$ branch of the dispersion respectively. The sub index $C$ means that these amplitudes are collected from a control system in which the saturation is constant throughout the sample. We repeat the process for every width, $L$, for values ranging between $4$ and $200$ nm.

\section{Results and Discussion} 
\begin{figure}
	\includegraphics[width=3.4in]{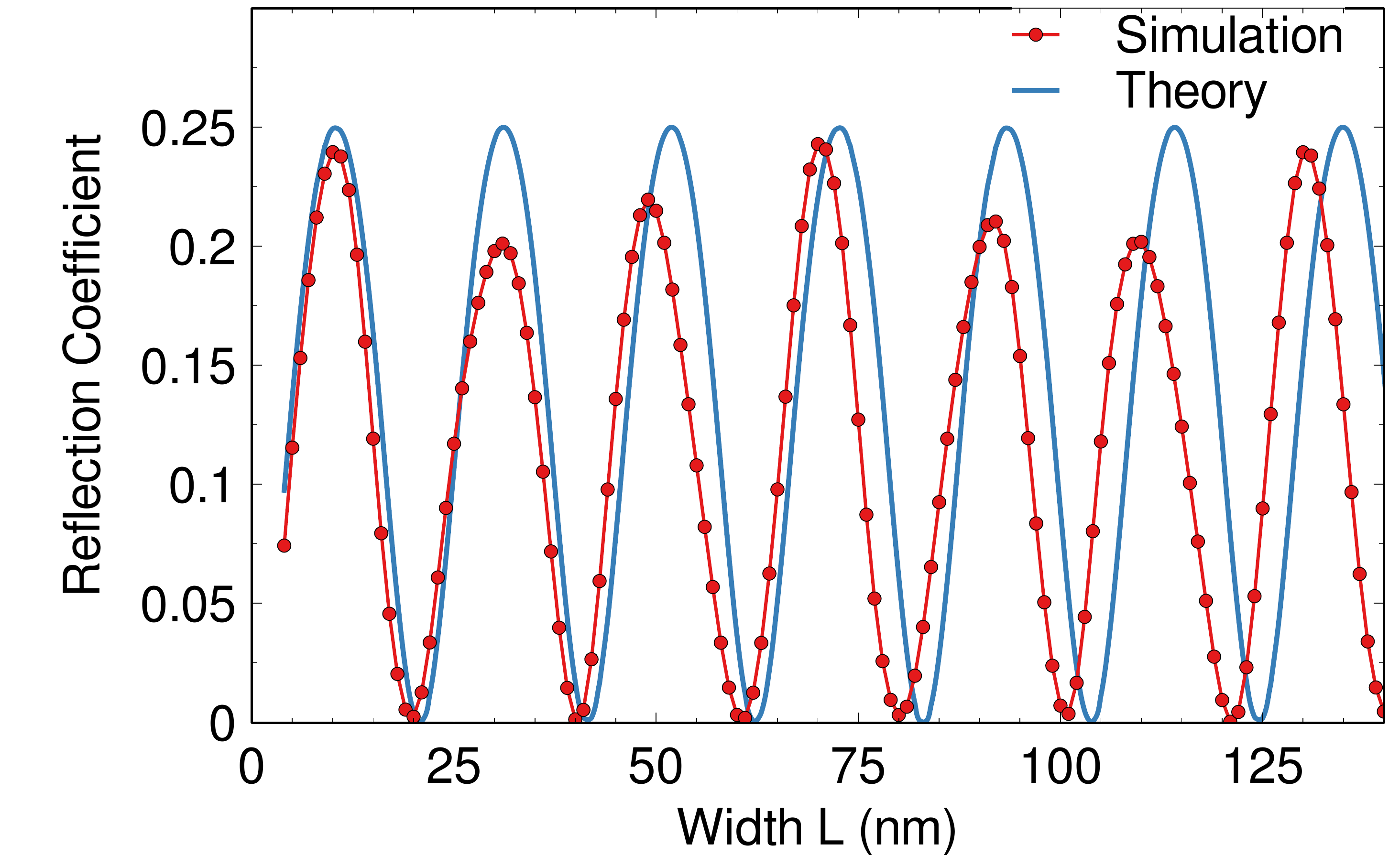}
	\caption{ Reflection coefficient as a function of the width, $L$, of Region II. The red, dotted curve shows the results from simulation, while the blue, solid line is obtained from eqn (\ref{eq:theoryRef}). A periodic behavior is shown characteristic for transmission resonances ($R=0$). }
	\label{fig:RvsL}		
\end{figure}

Fig (\ref{fig:RvsL}) shows the reflection coefficient as a function of  region II width, $L$. There are certain widths for which the reflection is minimum  consistent with resonant behavior. To analytically calculate the reflection coefficient, we consider that the components of the magnetization, $m_y$ and $m_z$, are plane wave-like solutions to the Landau-Lifshitz equation:

\begin{align}
	m_{y,z}&=Ae^{i(kx-\omega t)}+Be^{i(-kx-\omega t)}\;\;\;\;\;\;\;\;\;\text{Region\;I}\;(x<0)\nonumber\\
	m_{y,z}&=Fe^{i(kx-\omega t)}\;\;\;\;\;\;\;\;\;\;\;\;\;\;\;\;\;\;\;\;\;\;\;\;\;\;\;\;\;\text{Region\;III}\;(x>L),
\end{align}

where in Region I the waves with coefficients $A$ and $B$ describe the incident and reflected waves respectively. In Region III we only consider a transmitted wave with coefficient $F$. The angular frequency, $\omega=2\pi\nu$, and the wavevector, $k$, follow the spin wave dispersion relation,

\begin{align}\label{eq:disprel}
	\omega(k)=\sqrt{(\omega_k+\omega_H)(\omega_k+\omega_H+\omega_M)},
\end{align}

appropriate for the geometry under consideration. Here, $\omega_k=2\gamma Ak^2/M_s$, results from the exchange interaction, and  $\omega_H=\gamma \mu_0 H_0$ and  $\omega_M=\gamma \mu_0 M_s$ derive from Zeeman and demagnetizing terms respectively.

Inside region II, the magnetization components are

\begin{align}	
	m_{y,z}&=Ce^{i(-k'x-\omega t)}+De^{i(k'x-\omega t)}
	\;\;\;\;\;\;\;\nonumber\\&\;\text{Region\;II}\;(0<x<L)
\end{align}

where $k'\neq k$ is the wavevector inside region II and follows eq (\ref{eq:disprel}) changing $M_s$ by $M_s'$.\\

As usual, the spin waves need to  join smoothly at $x=0$, and $x=L$. Mathematically, the  boundary conditions are, 

\begin{align}\label{eq:BC}
	A+B&=C+D\nonumber\\
	kA-kB&=k'D-k'C\nonumber\\
	Ce^{-ik'L}+De^{ik'L}&=Fe^{-ikL}\nonumber\\
	k'De^{ik'L}-k'Ce^{-ik'L}&=kFCe^{ikL}
\end{align}  

and express the continuity of $m_{x,z}$, and $\partial m_{x,z}/\partial x$ both at $x=0$, and $x=L$. After some algebra, it is possible to find the reflection coefficient

\begin{align}\label{eq:theoryRef}
	R=\left|\frac{B}{A}\right|^2=\frac{\alpha^2}{1+\alpha^2},
\end{align}

with 

\begin{align}\label{eq:alpha}
	\alpha=\left[\frac{(k^2-k'^2)}{(2kk')}\right]\sin (k'L).
\end{align}

The reflection minima occurs when $\alpha=0$, i.e. when  $k'L=n\pi$, with $n=1,2,3,\dots$. The reflected wave is a superposition from waves reflected from the leading and trailing edges of region II at $x=0$, $x=L$, respectively. If these reflected waves arrive phase shifted by odd multiples of $\pi$, they interfere destructively leading to a perfect transmission ($R=0$). The wave reflected at $x=L$ travels an extra distance of $2L$ before superposing with the wave reflected at $x=0$, this results in a phase difference of $k'(2L)$. Also, there is an intrinsic phase shift of $\pi$ familiar even in classical waves. A traveling wave arriving at the interface separating two media is partially transmitted and partially reflected. The reflected portion is phase shifted by $\pi$ only in the case where the wave speed is lower in the medium being penetrated. Then, the destructive interference condition is $k'(2L)+\pi=(2n+1)\pi$ or 

\begin{align}\label{eq:CondInterference}
	k'L=n\pi
\end{align}

Eq (\ref{eq:theoryRef}) is plotted as a blue curve in fig (\ref{fig:RvsL}) for comparison with the results of the simulation (red curve). There is a good agreement between theory and simulation in the points where reflection is minimum for $L<60$ nm. For $L>60$ nm, simulation and theory start to phase out with simulation underestimating the transmission resonances. Regarding the value of the reflection coefficient, there is a reasonable agreement considering that the analytical calculations do not account for ellipticity.

\begin{figure}
	\includegraphics[width=3.4in]{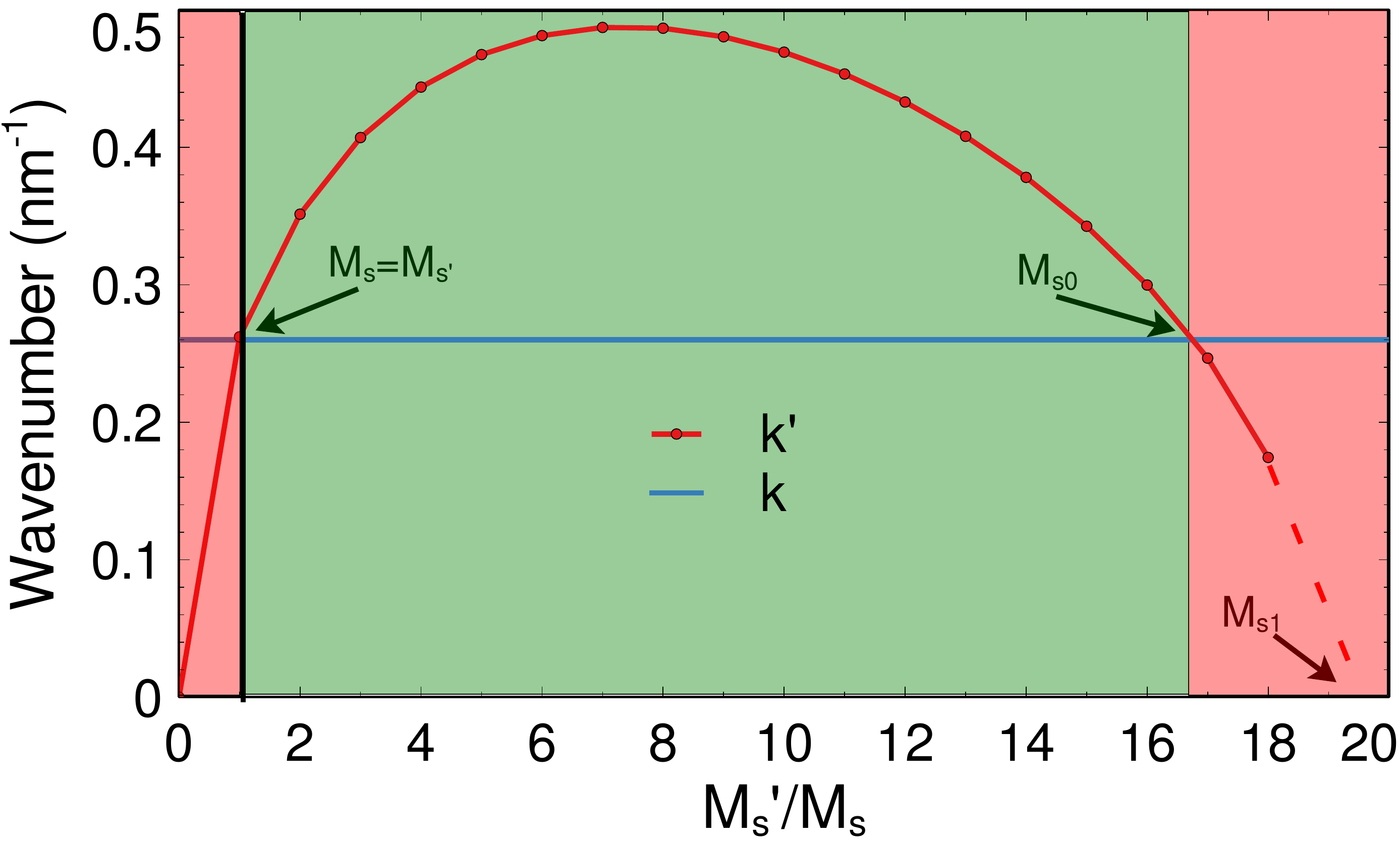}
	\caption{Comparison between wave numbers, $k$ and $k'$ when the saturation, $M_{s}'$, in Region II is varied.   }	
	\label{fig:PhaseShift}	
\end{figure}

To derive the interference condition, eq (\ref{eq:CondInterference}), it is only necessary to know that the intrinsic phase shift indeed happens. However, it may be relevant for future possible applications to determine where the intrinsic phase shift occurs ($x=0$ or $x=L$). To investigate this further, we rewrite dispersion relation, eq (\ref{eq:disprel}), as

\begin{align}\label{eq:kvsMs}
	k^2=2m^{*}(e-u).
\end{align}	

Where we have defined 

\begin{align}\label{eq:mass}
	m^*\equiv&\frac{M_s}{4\gamma A}\nonumber\\
	e\equiv&\sqrt{\omega^2-\left(\frac{\omega_M}{2}\right)^2}\nonumber\\
	u\equiv&\omega_H+\frac{\omega_M}{2}.
\end{align}

Note that equation (\ref{eq:kvsMs}) has the same mathematical form as kinetic energy, $p^2/2m=(\hbar k)^2/2m$, of a matter wave with energy, $E$, propagating above a potential well or barrier of magnitude $U$. From equation (\ref{eq:qmdisp})

\begin{align}\label{eq:momenta}
	p^2=2m(E-U).
\end{align}

Indeed, in equation (\ref{eq:mass}), $m^*$ corresponds to what is defined as the magnon mass; $e$ is the energy of the incident spin wave, $\hbar\omega$, modified by $\omega_M/2$; and $u$ is the magnitude of the potential energy as can be seen from  equation (\ref{eq:swdisp}). All terms divided by $\hbar$.

To determine whether the intrinsic phase shift occurs at $x=0$ or $x=L$ it suffices to compare the wave number in region II, $k'$ (given by eq (\ref{eq:kvsMs}) when $M_s$ is replaced by $M_s'$) with the wave number, $k$, in regions I and III (given by eq (\ref{eq:kvsMs}) with $M_s=300$ kA/m). As the wave speed, $v_w$, is inversely proportional to the wave number,$v_w=\omega/k$, when $k<k'$ the phase shift occurs at $x=0$, and when $k>k'$ the phase shift occurs at $x=L$. In figure (\ref{fig:PhaseShift}), we plot equation (\ref{eq:kvsMs}) as a function of $M_{s'}$ and obtain the red curve. This corresponds to theoretically varying the saturation magnetization in Region II. We also plot equation (\ref{eq:kvsMs}) for $M_s=300$ kA/m corresponding to the saturation magnetization in Regions I and III as the blue horizontal line. In this way we can compare the spin wave wavenumbers as they propagate through Region II with varying saturation magnetization. In the left red region, where $M_{s'}<M_s$ corresponding to a potential well, $k'<k$ and hence the phase shift is at $x=L$. Then there is a green region in the middle, $M_s<M_{s'}<M_{s0}\approx5100$ kA/m, which corresponds to a potential barrier, and where $k'>k$. In this section the phase shift is at $x=0$. While the right red region is still a potential well ($M_{s'}>M_s$), now $k'<k$ again and therefore the phase shift is at $x=L$. Finally,  $M_{s1}$ is given by the value where $k'$ becomes imaginary (i.e. $e'<u'$). Above $M_{s1}\approx5700$ (kA/m) tunneling should be observed.  

It is worth noting that at $M_{s0}$, $k'=k$ and hence $\alpha=0$ in eq (\ref{eq:alpha}), so that the spin wave is perfectly transmitted. This situation is reminiscent to exchange spin waves propagating through a domain wall without reflection. The domain wall, described by the P\"oschl-Teller potential, has the \textit{exact depth} so that the potential is reflectionless~\cite{Vasiliev,lekner_reflectionless_2007}. However, when the domain wall is disturbed (e.g. through the DMI) reflection is found~\cite{borys_spinwave_2016}. In our case, $M_{s0}$ has the \textit{exact height} to be invisible for the spin waves. While the values $M_{s0}$ and $M_{s1}$ are unrealistically high and could never be obtained through a thermal landscape, the reflection coefficient in figure (\ref{fig:RvsL}) (with $M_{s}'=100$ kA/m) is experimentally achievable. This discussion is presented for a complete analysis of eq \ref{eq:alpha}, and reference for future studies in different (larger or perpendicularly magnetized) systems.   

The variation of the wave number $k'$ as a function of the saturation, is distinctly different from how the momentum, $p=\hbar k$, varies as a function of a potential field in the quantum mechanical case. Mainly, the magnon mass, $m^*$, is modified when the spin wave goes from one region to the next. Also, the modified energy of the incident spin wave, $e$, depends on the saturation and therefore is different in each region.  

\begin{figure}
	\includegraphics[width=3.4in]{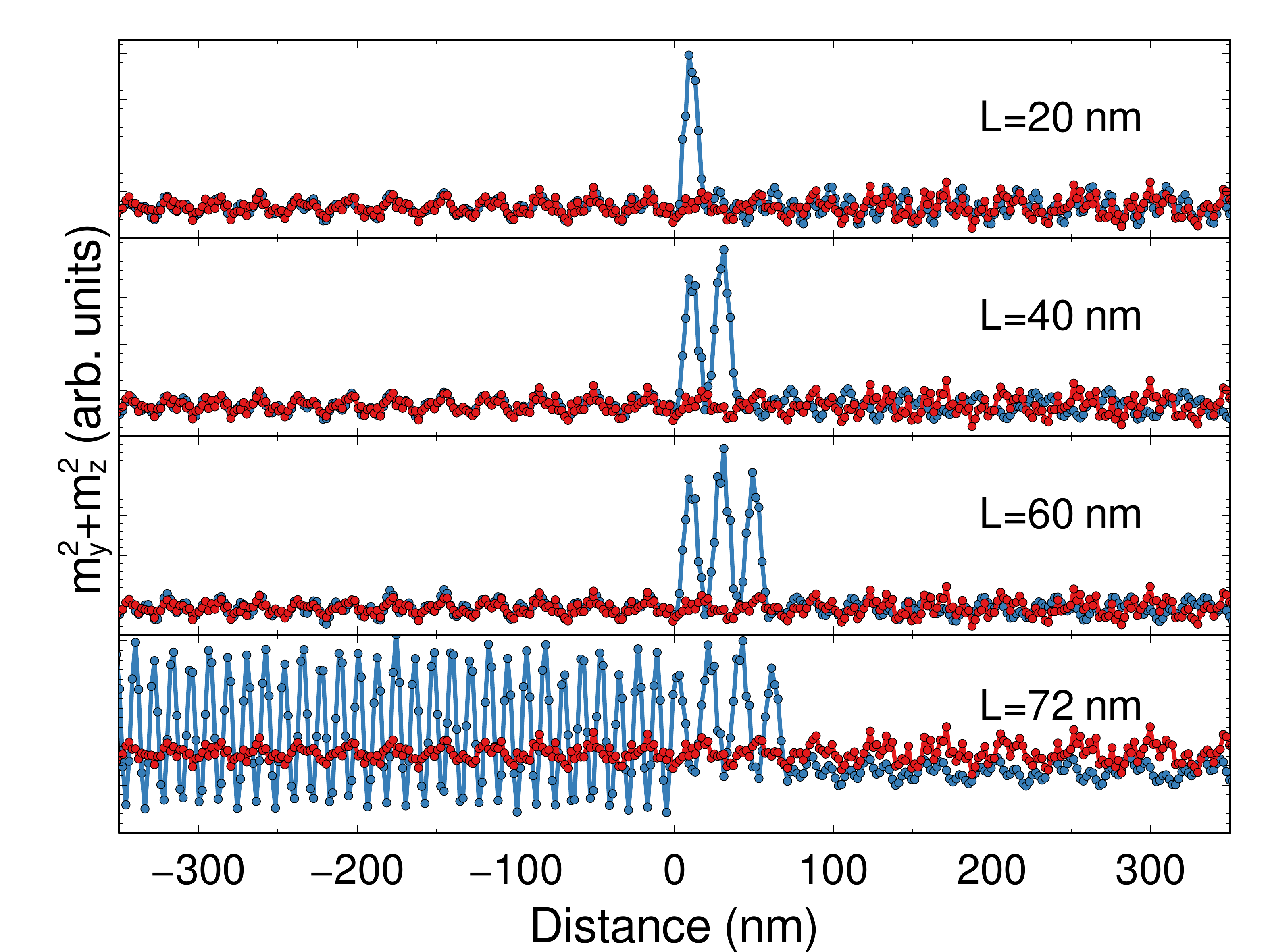}
	\caption{ Squared amplitudes $m_y^2+m_z^2$ for $L=20$, $40$, $60$, and $72$ nm in blue. In red, the squared amplitudes of a homogeneous system for comparison. In the top three panels, which correspond to widths where reflection is minima, the squared amplitude \textit{fits} an integer number of times within Region II. In the bottom panel, squared amplitude does not fit an integer number of times within the well and hence reflection is maximum.   }	
	\label{fig:amplitude}	
\end{figure}

While there are some differences between the quantum mechanical case and the spin wave case, the underlying physical process is the same. In figure (\ref{fig:amplitude}) we plot the squared amplitude $m_y^2+m_z^2$ for widths in which the reflection is minimum. It can be seen that for these widths, the squared amplitude  \textit{fits} an integer number within the well. This phenomenon resembles what happens in the quantum mechanical case in which the transmission resonances occur whenever the probability density $|\psi|^2$ fits in the well an integer number of times.

The interference condition, $k'L=n\pi$ can be written in terms of modified incident energy, $e$,  of the spin wave as (compare to eq (\ref{eq:RTresonances}))

\begin{align}\label{eq:ResonancesEn}
	e'=u'+\frac{n^2\pi^2}{2m'^*L^2},\;\;\;\;\;\;\;\;\;\;\;\;\;\;n=1,2,3,\dots.
\end{align}

\begin{figure}
	\includegraphics[width=3.4in]{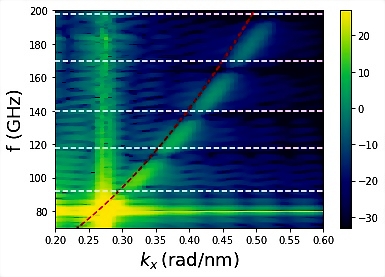}
	\caption{ Amplitudes of the $y$ component of the magnetization in frequency-reciprocal space diagram. Black horizontal regions for certain frequencies match the frequencies (white dashed lines) obtained from eqn (\ref{eq:ResonancesEn}).The red, dashed curve is the dispersion relation for comparison.  }	
	\label{fig:RvsE}	
\end{figure}

The energies for which transmission resonances  vary quadratically as a function of number $n$ and has the same mathematical form as its quantum mechanical counterpart. 
For fixed $L$, and $M_s'=100$ (kA/m), there are a set of values of $e$ where the reflection is minimum. We observe the case $L=108$ nm which corresponds neither to a maximum nor a minimum in $R(L)$. Simulation results are shown in figure (\ref{fig:RvsE}). Black horizontal lines can be distinguished in the yellow, dispersion curve corresponding to the frequencies where the resonance condition holds. The white horizontal dashed lines are the theoretical values calculated from equation (\ref{eq:ResonancesEn}) for $n=1,2,3,4$.  The minima in the reflection match well with the theoretical values. 

\section{Conclusions}

We showed theoretically and with simulations that spin waves exhibit the quantum mechanical RT effect where  there are transmission resonances ($R=0$). We found that such resonances occur for certain widths of a  potential well or for certain energies if the width of the well is fixed. For a fixed width it is necessary a substantial increase in energy to go from one resonance to the next. Also, it is not possible to dynamically control the system as the excitation energy needs to be changed for each resonance. On the other hand, it is possible to control the width of the well by creating a suitable temperature landscape resulting in a real dynamical, reconfigurable system. 	
We determined whether the intrinsic phase shift a wave suffers when its velocity decreases as it penetrates a different medium occurs at $x=0$, or at $x=L$ as a dependence of $M_s'$. For manipulation of data this information may be useful as logic computation can be achieved.  We also found that there is a certain value of the saturation, $M_{s0}$, which makes a potential barrier invisible for the spin waves. The situation resemble what happens with exchange spin waves propagating through a domain wall with no reflection but with a phase shift. Determining if there is a phase shift for perfectly transmitted spin waves would be useful for logic operations.	
Our findings further contribute to make devices based in reconfigurable magnetic materials more efficient. For example, by determining the widths, energies, and saturation where reflection is minimum or maximum it is possible to control the band gap in a magnonic crystal. Moreover, full understanding of how spin waves behave in devices where saturation is modulated may lead to an alternative to the race track memory device. If regions with different saturation describe different states, it may be plausible to move the region (hence the state) by changing the thermal landscape. 	
It is necessary, however, to determine how increasing the dimensions of the sample affects our results. Because another quantum mechanical effect, tunneling, has been observed in larger samples, we expect that as long as the thickness of the sample remains less than tens of nanometers, demagnetizing effects will not affect the results. 

For a material such as YIG, it would be possible to get a saturation magnetization of $100$ kA/m using a thermal landscape with a temperature of $425$ K, well below its Curie temperature ($559$ K). For other magnetic materials a thermal landacape could not result in the necessary modulation of the saturation magetization. In such cases, however, ion implantation could be an alternative. For example, in CoCrPt \cite{marko2010determination} and CoFe \cite{mcgrouther2005nanopatterning} films, saturation magnetization was totally suppressed using FIB irradiation or Cr implantation into Permalloy film to locally reduce the Curie temperature\cite{fassbender2008magnetic}. Modifying saturation magnetization with ion implantation would mean that the desirable dynamic control attribute is lost. In any case, the periodical behavior of the reflection should remain.

\acknowledgments
The authors acknowledge fruitful discussions with A.A.
Serga and G. Venkat. This work was partially supported by the National
Council of Science and Technology of Mexico (CONACyT) under project Fronteras No. 344 and CB 253754, UNAM-DGAPA IN107318, and by fellowship
Beca UNAM postdoctoral.

\bibliography{articles}
\bibliographystyle{eplbib}

\end{document}